\documentclass[aps,prl,twocolumn,superscriptaddress,amsmath,amssymb,showpacs]{revtex4}

\usepackage{graphicx}
\usepackage{dcolumn}
\usepackage{bm}
\usepackage{color}

\begin{document}

\title{A new superconductor derived from topological 
insulator heterostructure}

\author{Satoshi Sasaki}
\author{Kouji Segawa}
\author{Yoichi Ando}
\email{y_ando@sanken.osaka-u.ac.jp}
\affiliation{Institute of Scientific and Industrial Research,
Osaka University, Ibaraki, Osaka 567-0047, Japan}

\date{\today}

\begin{abstract}

Topological superconductors (TSCs) are of significant current interest
because they offer promising platforms for finding Majorana fermions.
Here we report a new superconductor synthesized by intercalating Cu into
a naturally-formed topological insulator (TI) heterostructure consisting
of Bi$_2$Se$_3$ TI units separated by nontopological PbSe units. For the
first time in a TI-based superconductor, the specific-heat behavior of
this material suggests the occurrence of unconventional
superconductivity with gap nodes. The existence of gap nodes in a
strongly spin-orbit coupled superconductor would give rise to spin-split
Andreev bound states that are the hallmark of topological
superconductivity. Hence, this new superconductor emerges as an
intriguing candidate TSC.

\end{abstract}





\maketitle

A major theme in current condensed matter physics is to understand and
explore the roles of topology in quantum mechanics. In topological
insulators (TIs), a nontrivial topology of the quantum-mechanical wave
functions leads to the appearance of gapless conducting states at the
boundary (i.e. edge or surface) \cite{QiZhang, HasanKane, Ando}.
Topological superconductors (TSCs) are conceptually similar to TIs and
are characterized by gapless quasiparticle states at the boundary
\cite{QiZhang, Schnyder08, Tanaka12, Maeno}, but an important
distinction from TIs is that the boundary state of a TSC is a good place
to look for Majorana fermions \cite{Alicea, Beenakker, Fu10}, which
possess a distinct property that particles are their own antiparticles
and would be useful for fault-tolerant topological quantum computing. In
this context, superconductors derived from TIs are of particular
interest, because the strong spin-orbit coupling inherent to TIs may
lead to unconventional pairing that is prerequisite to TSCs \cite{Fu10}.
However, there have been only a few cases in which superconductivity is
found in doped TIs \cite{Hor, Sasaki, Sasaki2, Bay, Paglione, LaPtBi,
YPtBi}, and it is strongly desired that a new superconductor with
promising indications of unconventional superconductivity is discovered
in a doped TI.

Recently, two of the authors have contributed to the discovery of an
interesting new topological insulator \cite{Nakayama},
(PbSe)$_5$(Bi$_2$Se$_3$)$_{6}$. This is a member of the Pb-based
homologous series of compounds \cite{Fang},
(PbSe)$_5$(Bi$_2$Se$_3$)$_{3m}$ ($m = 1,2,\cdots$), which naturally form
multilayer heterostructures of a topological insulator (Bi$_2$Se$_3$)
and an ordinary insulator (PbSe). It was found that at $m$ = 2, the PbSe
unit works as a block layer and the topological boundary states are
encapsulated in each Bi$_2$Se$_3$ unit, making the system to possess
quasi-two-dimensional (quasi-2D) states of topological origin throughout
the bulk. Bi$_2$Se$_3$ consists of a stack of Se-Bi-Se-Bi-Se quintuple
layers (QLs), and the $m$ = 2 member of (PbSe)$_5$(Bi$_2$Se$_3$)$_{3m}$
has 2 QLs in its Bi$_2$Se$_3$ unit [see Fig. 1(a)]. In the middle of this
2-QL unit is a van der Waals gap, into which intercalations of atoms or
molecules are possible. 

Motivated by the occurrence of superconductivity in Bi$_2$Se$_3$ upon Cu
intercalation \cite{Hor}, we tried to make
(PbSe)$_5$(Bi$_2$Se$_3$)$_{6}$ (hereafter called PSBS) superconducting
via Cu intercalation. We adopted the electrochemical technique which we
developed for making high-quality Cu$_x$Bi$_2$Se$_3$ superconductors
\cite{Kriener11_1}, and we have succeeded in synthesizing \cite{SM} a
new superconductor with this strategy. Intriguingly, this new material,
Cu$_x$(PbSe)$_5$(Bi$_2$Se$_3$)$_6$ (called CPSBS), turned out to be
quite different from its cousin, Cu$_x$Bi$_2$Se$_3$: First, this new
superconductor presents an unusual specific-heat behavior which suggests
unconventional superconductivity. Second, nearly 100\% superconducting
samples can sometimes be synthesized, which makes it easier to elucidate
its intrinsic nature.

\begin{figure}
\includegraphics*[width=8.7cm]{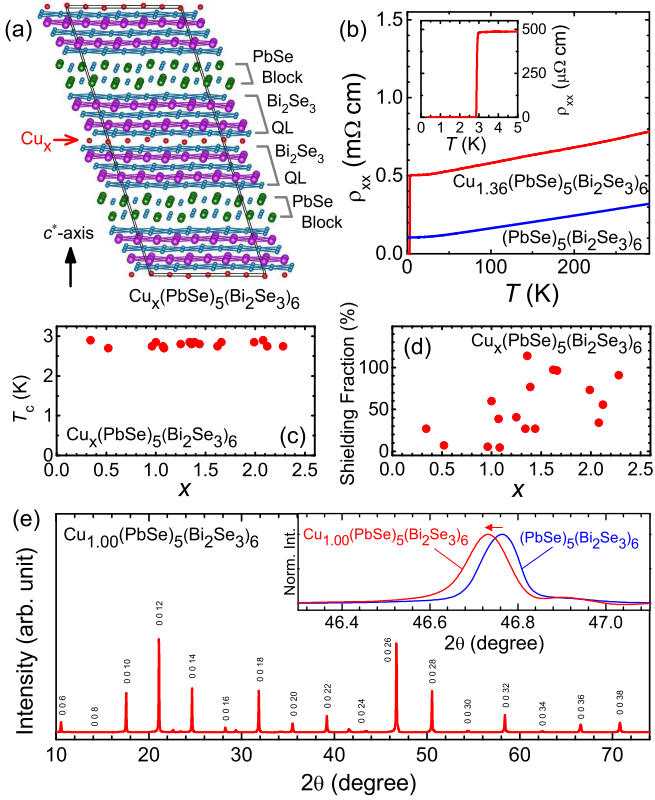}
\caption{(Color online) 
Cu$_x$(PbSe)$_5$(Bi$_2$Se$_3$)$_6$ superconductor.
(a) Crystal structure based on the data for PSBS
\cite{Fang}. Cu atoms are intercalated into the van der Waals gaps
marked by red arrow.
(b) Temperature dependencies of the resistivity $\rho_{xx}$ for
PSBS and CPSBS ($x$ = 1.36). Inset shows the sharp superconducting
transition at 2.85 K.
(c) Onset $T_c$ measured by dc magnetic
susceptibility for samples with various $x$ values.
(d) Shielding fractions of the samples presented in (c) at 1.8 K.
(e) XRD pattern measured on a cleaved surface ($ab$ plane) of a
superconducting CPSBS sample with $x$ = 1.00 (shielding fraction
$\sim$60\%), where the peak pattern is essentially the same as that of
pristine PSBS. Inset compares the positions of the prominent (0 0 26)
peak for PSBS and CPSBS, which demonstrates that the periodicity
perpendicular to the layers is slightly enlarged from 50.460(1) \AA\ to
50.508(1) \AA\ after the Cu intercalation.
} 
\label{Fig1}
\end{figure}

Figure 1(b) shows the resistivity data for PSBS and CPSBS ($x$ = 1.36).
The Cu intercalation causes a sharp superconducting transition at 2.85
K, and at the same time, it introduces moderate electron scattering to
enhance the residual resistivity. The carrier density increases from
$n_e \simeq 4 \times 10^{20}$ in PSBS to $1.2 \times 10^{21}$ cm$^{-3}$
in CPSBS ($x$ = 1.36), which suggests that each intercalated Cu
introduces about 0.7 electron on average \cite{SM}. The Hall resistivity
data indicate that the transport is governed by only one band (see Fig.
S1(b) of the Supplemental Material \cite{SM}), suggesting that the
topological and nontopological bands of the Bi$_2$Se$_3$ unit
\cite{Nakayama} may well have merged at the chemical potential of CPSBS.
As shown in Figs. 1(c) and 1(d), in the present set of samples the onset
of superconductivity was essentially independent of $x$ and was always
around 2.9 K for $x$ = 0.3 -- 2.3, whereas the shielding fraction (the
fraction of the sample volume from which the magnetic field is kept out
due to superconductivity after zero-field cooling) was very much sample
dependent; note that, in the case of type-II superconductors, the
shielding fraction is a better measure of the superconducting volume
fraction than the Meissner fraction measured upon field cooling, because
the latter is significantly affected by flux trapping. The random nature
of the obtained shielding fraction vs $x$ signifies the difficulty in
synthesizing a homogeneous superconductor with intercalation, and a
majority of our samples are inhomogeneous. Nonetheless, we have been
able to achieve essentially 100\% shielding fraction in a few samples
with $x$ = 1.3 -- 1.7, and in those special samples the roles of
minority phases can be largely neglected. The x-ray diffraction (XRD)
data from cleaved surfaces of single-crystalline CPSBS indicate that the
system essentially preserves the same crystal structure of PSBS with a
slightly elongated $c^*$-axis, as is expected for an intercalated
material [Fig. 1(e)]; however, it is beyond the scope of this paper to
precisely determine the crystallographic structure, including the exact
position of Cu, of this obviously complicated material.

\begin{figure}
\includegraphics*[width=8.7cm]{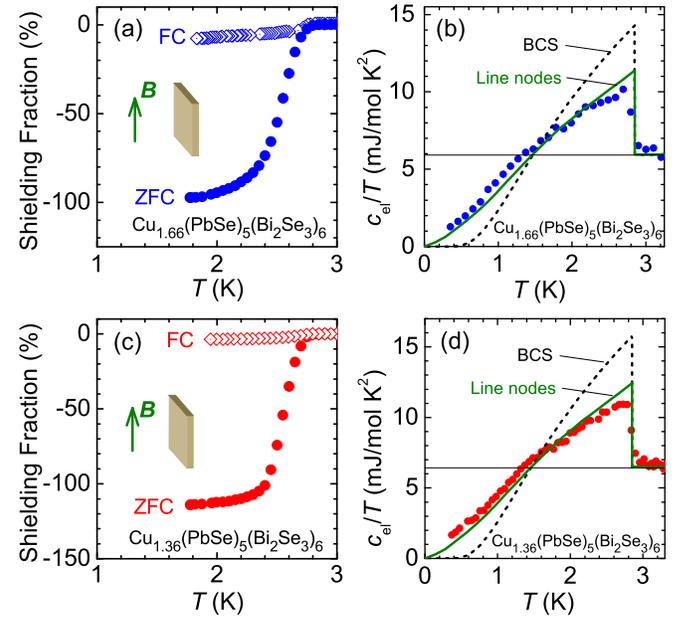}
\caption{(Color online) 
Shielding fraction and specific heat.
(a) and (c) Temperature dependencies of the dc magnetic
susceptibility measured in 0.2 mT applied parallel to the $ab$ plane
with the field-cooling (FC) and zero-field-cooling (ZFC) procedures for
(a) $x$ = 1.66 and (c) $x$ = 1.36, presented in terms of the shielding
fraction. Since the demagnetization effect is minimal for this geometry
and the sample shape was irregular,
we did not make any correction for it [the $x$ = 1.66 (1.36) sample was
0.23 (0.16) mm thick and 1.6 $\pm$ 0.3 (1.7 $\pm$ 0.35) mm long along 
the $B$ field].
(b) and (d) Superconducting transition in $c_{\rm el}/T$ in
0 T obtained after subtracting the phonon contribution determined in 2 T
(see Fig. S2 \cite{SM}). The dashed line is 
the weak-coupling BCS behavior (coupling constant $\alpha$ = 1.76) for 
$T_c$ of 2.85 K. The green solid line is the theoretical curve for 
$d$-wave pairing on a simple cylindrical Fermi surface with line nodes 
along the axial direction \cite{Maki}. Horizontal solid line corresponds 
to $\gamma_{\rm N}$.
} 
\label{Fig2}
\end{figure}

In the following, we focus on two samples with $x$ = 1.36 and 1.66,
which presented essentially 100\% shielding fractions as shown in Figs.
2(a) and 2(c). Figures 2(b) and 2(d) show the behavior of the electronic
specific heat $c_{\rm el}$ in terms of $c_{\rm el}/T$ vs $T$ for the two
samples; those data were obtained after subtracting the phonon
contribution determined in 2 T described in Fig. S2 of the Supplemental 
Material \cite{SM}. The two samples consistently present two unconventional 
features that become apparent when compared with the conventional 
weak-coupling BCS behavior \cite{Tinkham} shown with dashed lines: 
First, the jump height at $T_c$ is much smaller than the prediction 
of the BCS theory, $1.43\,\gamma_{\rm N}$, where $\gamma_{\rm N}$ is 
the normal-state electronic specific-heat coefficient corresponding to 
the horizontal solid lines. Second, $c_{\rm el}/T$ decreases much more 
slowly than the BCS behavior; in particular, $c_{\rm el}/T$ keeps showing 
a sizable temperature dependence even at our lowest temperature of 0.35 K 
($T/T_c$ = 0.12), whereas $c_{\rm el}/T$ should already become negligible 
at such a low temperature in the BCS case. It is reassuring that those
unconventional features are exactly reproduced in two different samples.

Such a peculiar behavior in $c_{\rm el}/T$ suggests the existence of
nodes in the superconducting gap for the following reasons: First, when
the gap has nodes, the averaged gap magnitude becomes smaller than the
fully-gapped case, and the specific-heat jump is naturally reduced
\cite{Maeno, Tinkham}; the green solid line in Figs. 2(b) and 2(d) gives an
example for the $d$-wave superconductivity with line nodes \cite{Maki}.
Second, in contrast to the conventional BCS case in which $c_{\rm el}/T$
decreases exponentially at low $T$ because of a finite activation
energy, the existence of nodes allows thermal excitations of
quasiparticles down to very low temperatures, changing the $T$
dependence of $c_{\rm el}/T$ from exponential to a power law
\cite{Maeno}. 

As one can see in Figs. 2(b) and 2(d), our data, particularly the strong
$T$ dependence near 0 K, bear striking similarity to the theoretical
$c_{\rm el}/T$ behavior expected for a superconductor with line nodes
\cite{Maki}, which point to the realization of unconventional
superconductivity in CPSBS. Of course, specific-heat measurements alone
are not sufficient for unambiguously nailing down the existence of
nodes, because a multiband superconductor with a very small gap in one
of the bands or an anisotropic $s$-wave superconductor with very small
gap minima would give rise to a $c_{\rm el}/T$ behavior similar to what
we found in CPSBS. Hence, phase-sensitive measurements are crucially
important in the future research of this material. Also, STM and NMR
experiments would be very useful for elucidating the realization of
unconventional superconductivity.

\begin{figure}
\includegraphics*[width=8.7cm]{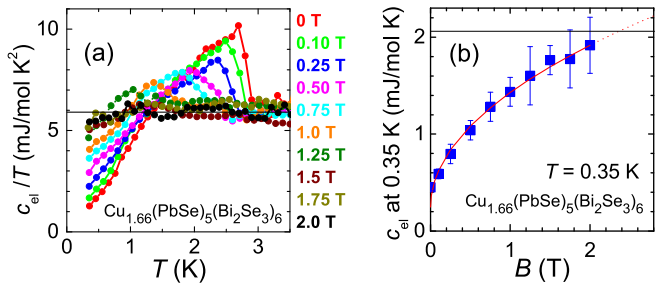}
\caption{(Color online) 
Specific heat in magnetic fields.
(a) Temperature dependencies of
$c_{\rm el}$/$T$ in various magnetic fields ($B$$\perp$$ab$) for $x$ =
1.66. Small Schottky anomaly that becomes non-negligible above $\sim$1.5 T
has been subtracted (see \cite{SM}). Horizontal solid line 
corresponds to $\gamma_{\rm N}$.
(b) Magnetic-field dependence of $c_{\rm el}$ at 0.35 K taken from 
the data in (a). The red solid line is the best fit of the function 
$aB^n + c_0$ to the data, yielding $n = 0.50 \pm 0.06$, $a = 1.2 \pm 0.1$ 
(mJ mol$^{-1}$ K$^{-1}$ T$^{-n}$), and $c_0 = 0.25 \pm 0.09$ 
(mJ mol$^{-1}$ K$^{-1}$). Horizontal solid line corresponds to 
$\gamma_{\rm N}T$ at $T$ = 0.35 K.
} 
\label{Fig3}
\end{figure}

Note that in the case of nodal superconductors, impurity scattering
causes a finite density of quasiparticle states at 0 K, causing $(c_{\rm
el}/T)_{T \rightarrow 0}$ to be finite even in a 100\% superconducting
sample; this may also be the case in the present system, given the
relatively large residual resistivity. Also, it is prudent to mention
that the spin-orbit scattering \cite{Maki-SO} is pair breaking and may
mimic the $c_{\rm el}/T$ behavior observed here, so its role should be
elucidated in future. Nevertheless, it is fair to note that in
Cu$_x$Bi$_2$Se$_3$, where the spin-orbit scattering should be of
similar strength, the $c_{\rm el}/T$ behavior was found to obey the
simple BCS theory \cite{Kriener11_1}.

Due to the quasi-2D nature of the parent material PSBS \cite{Nakayama},
the superconductivity in CPSBS is likely to be realized on a quasi-2D
Fermi surface, which is distinct from the three-dimensional (3D) bulk
Fermi surface of Bi$_2$Se$_3$. This implies that the theory of 3D
topological superconductivity proposed for Cu$_x$Bi$_2$Se$_3$
\cite{Fu10} is not directly applicable. Nevertheless, it is still
expected that strong spin-orbit coupling responsible for the topological
nature of the parent material causes the effective pairing interaction
to become spin dependent, which would lead to unconventional
superconductivity \cite{Fu10}. When the Fermi surface is quasi-2D, a
node in the unconventional superconducting gap is naturally extended
along the $c^*$ axis, forming a line node in the 3D Brillouin zone. It
is thus expected that, if gap nodes were to be present in CPSBS, the
$c_{\rm el}/T$ behavior should be close to that of a superconductor with
line nodes.

The possible existence of line nodes in CPSBS is further supported by
the magnetic-field dependence of the specific heat. Figure 3(a) shows the
$c_{\rm el}/T$ vs $T$ plots for various magnetic fields applied
perpendicular to the $ab$ plane described in Fig. S2 of the Supplemental 
Material \cite{SM}, from which we extract the magnetic-field dependence 
of $c_{\rm el}$ at the lowest temperature, 0.35 K; here, to make our best 
effort to quantify its behavior, the data are corrected for a small 
Schottky anomaly \cite{SM}, which is only $\lesssim$ 20\% even at the 
upper critical field and is comparable to the error bar. The obtained 
$c_{\rm el}(B)$ behavior at 0.35 K [Fig. 3(b)] is clearly nonlinear in $B$. 
Note that in conventional BCS superconductors $c_{\rm el}$ increases 
essentially linearly with $B$, because the number of induced quasiparticles 
is proportional to the number of vortices. On the other hand, in nodal
superconductors, the Doppler shift of the quasiparticle excitations
(so-called Volovik effect) causes more quasiparticles to be created per
vortex than in the BCS case \cite{Volovik}; for line nodes, Volovik
showed \cite{Volovik} that $c_{\rm el}$ increases as $\sim$ $\sqrt{B}$.
Indeed, our data are best described with $c_{\rm el} \sim B^{0.5}$,
supporting the existence of line nodes.

\begin{figure}
\includegraphics*[width=8.7cm]{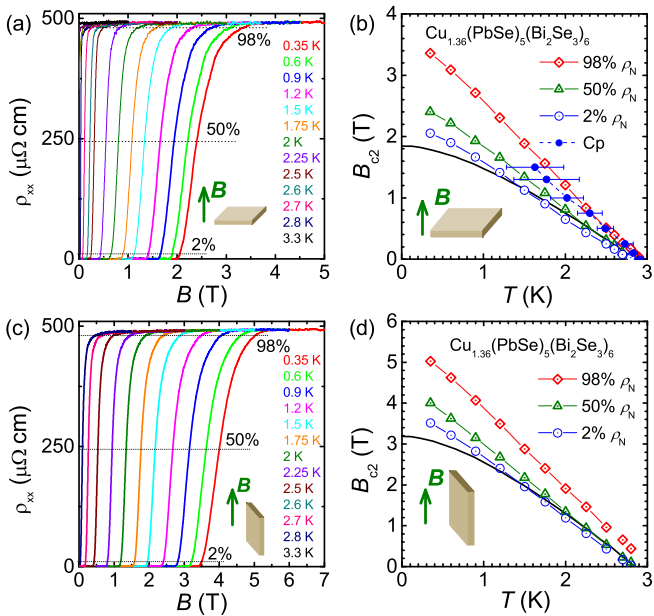}
\caption{(Color online) 
Upper critical field $B_{c2}$.
(a) and (c) Magnetic-field-induced resistive transitions
measured at various temperatures in the $x$ = 1.36 sample for (a)
$B$$\perp$$ab$ and (c) $B$$\parallel$$ab$. Three levels of the
resistivity $\rho_{xx}$ in those transitions, corresponding to 2\%,
50\%, and 98\% of $\rho_{\rm N}$ (shown by dotted lines), are used for
determining the depinning, mid-point, and onset fields, respectively;
50\% $\rho_{\rm N}$ is the definition of $B_{c2}$.
(b) and (d) $B$ vs $T$ phase diagrams obtained from the data
in (a) and (c), respectively. The black solid lines are the conventional
$B_{c2}(T)$ behavior given by the WHH theory, which is determined 
by the $(dB_{c2}/dT)_{T=T_c}$ value (0.936 T/K for $B$$\perp$$ab$ and
1.615 T/K for $B$$\parallel$$ab$). In (b), the onset temperatures of the
specific-heat transitions in various magnetic fields measured in the
same sample (see \cite{SM}) are also shown for comparison.
} 
\label{Fig4}
\end{figure}

Figures 4(a) and 4(c) show the magnetic-field-induced resistive transitions
at various temperatures in $B$$\perp$$ab$ and in $B$$\parallel$$ab$,
respectively, from which the resistive upper critical fields $B_{c2
\perp}(T)$ and $B_{c2 \parallel}(T)$ are extracted. We plot in Figs. 4(b)
and 4(d) the magnetic field values at which 2\%, 50\%, and 98\% of the
normal-state resistivity $\rho_{\rm N}$ is recovered at a given
temperature. It is customary to use 50\% $\rho_{\rm N}$ for defining
$B_{c2}$ \cite{Bay, Paglione, Maple}. In Fig. 4(b), the onset temperatures
of the specific-heat transitions in $B$$\perp$$ab$ determined for the
same sample (see Fig. S5(a) of the Supplemental Material \cite{SM}) 
is also shown. Extrapolations of the 50\% $\rho_{\rm N}$ data in 
Figs. 4(b) and 4(d) give $B_{c2 \perp}$ = 2.6 T and 
$B_{c2 \parallel}$ = 4.3 T at 0 K, yielding the coherence lengths 
$\xi_{ab}$ = 11.3 nm and $\xi_{c^*}$ = 6.8 nm. The relatively small 
anisotropy in $B_{c2}$ may seem strange for a superconductor with
a quasi-2D Fermi surface, but a similar situation has been found in
BaFe$_2$As$_2$-based superconductors \cite{BKFA, BFCA} and is believed
to be due to a finite $k_z$ dispersion of the cylindrical Fermi surface.

The $B_{c2}(T)$ behavior expected for a conventional superconductor from
the Werthamer-Helfand-Hohenberg (WHH) theory \cite{WHH} tends to
saturate for $T \rightarrow 0$, as shown with solid lines in Figs. 4(b)
and 4(d). On the other hand, our experimental data present much weaker
tendency toward saturation. We note that the Pauli paramagnetic limit
\cite{WHH}, $B_{\rm Pauli} = 1.84\, T_c$ = 5.3 T, is larger than our
$B_{c2}$, so the violation of the conventional behavior is not as strong
as in the case of exotic superconductors like UBe$_{13}$ \cite{Maple}.
Nevertheless, similar violations of the WHH theory as is found here have
been discussed to be indicative of unconventional superconductivity in
Cu$_x$Bi$_2$Se$_3$ \cite{Bay} and in Bi$_2$Se$_3$ under high pressure
\cite{Paglione}.

We have further characterized the CPSBS superconductor by measuring the
lower critical field $B_{c1}$, which was determined to be 0.34 mT for
$B$$\parallel$$ab$ at 0 K (Fig. S6(c) of the Supplemental Material \cite{SM}). 
Knowing $B_{c1 \parallel}$, $B_{c2\perp}$, and $B_{c2 \parallel}$, one can 
obtain the Ginzburg-Landau parameter $\kappa_{ab}$ = 192, the penetration 
depths $\lambda_{ab}$ = 1.3 $\mu$m and $\lambda_{c^*}$ = 2.2 $\mu$m, and 
the thermodynamic critical field $B_c$ = 16.6 mT \cite{SM}. 
The long penetration depths are consistent with the low carrier density and 
moderate disorder in CPSBS.

It is striking that all the bulk properties of the new superconductor
CPSBS shown here point to possible occurrence of unconventional
superconductivity accompanied with line nodes. The existence of line
nodes implies a sign-changing gap function, which generically gives rise
to surface Andreev bound states \cite{Tanaka}. Strong spin-orbit
coupling causes such surface Andreev states to be spin-split and form
spin-non-degenerate Kramers pairs, which means that they become helical
Majorana fermions \cite{Fu10}. Indeed, nodal superconductors with strong
spin-orbit coupling have been discussed to be topological
\cite{Sasaki,Sato10}. Therefore, the superconductivity in CPSBS has a
good chance to be topological and harbour Majorana fermions. To nail
down the topological nature, making a Josephson junction with a
conventional superconductor to measure a nontrivial current-phase
relationship coming from boundary Majorana fermions \cite{Alicea,
Beenakker} would be a smoking-gun experiment.

\begin{acknowledgments} 

We thank T. Toba for his help in synthesizing the samples, K. Eto and 
M. Kriener for technical assistance, and L. Fu, Y. Tanaka, A. Taskin, 
and A. Yamakage for helpful discussions. This work was supported by 
JSPS (KAKENHI 24740237, 24540320, and 25220708), MEXT (Innovative Area
``Topological Quantum Phenomena" KAKENHI), AFOSR (AOARD 124038), and
Inamori Foundation.

\end{acknowledgments}

\clearpage
\onecolumngrid

\renewcommand{\thefigure}{S\arabic{figure}} 

\setcounter{figure}{0}

\renewcommand{\thesection}{S\arabic{section}.} 

\begin{flushleft} 
{\Large {\bf Supplemental Material}}
\end{flushleft} 

\vspace{2mm}

\begin{flushleft} 
{\bf Materials and methods}
\end{flushleft} 

High-quality single crystals of (PbSe)$_5$(Bi$_2$Se$_3$)$_{6}$ (PSBS)
were grown by a modified Bridgman method using high purity elements Pb
(99.998\%), Bi (99.9999\%), and Se (99.999\%) with the starting
composition of Pb:Bi:Se = 7:26:46 in a sealed evacuated quartz tube at
698 $^{\circ}$C for 6 h, followed by a slow cooling to 650 $^{\circ}$C
with a cooling rate of 12 $^{\circ}$C/day and then quenching to room
temperature. The phase diagram of the Pb-Bi-Se ternary system is very
complicated \cite{Sherimova}, which causes multiple crystal phases to
coexist in a boule. After the growth, we chose the PSBS phase based on
the x-ray diffraction analysis of the crystals cut out from the boule
\cite{Nakayama}. Roughly 30\% of a boule is in the $m$ = 2 PSBS phase.
For the electrochemical Cu intercalation, we used a saturated solution
of CuI powder (99.99\%) in acetonitrile CH$_3$CN \cite{Kriener11_2}.
Samples with a typical size of $2 \times 1 \times 0.2$ mm$^3$ were wound
by a 50-$\mu $m thick, bare Cu wire, and they together acted as the
working electrode. A 0.5-mm thick Cu stick was used as both the counter
and reference electrode. The concentration of intercalated Cu was
determined from the weight change before and after the intercalation
process, giving the $x$ value of Cu$_x$(PbSe)$_5$(Bi$_2$Se$_3$)$_{6}$
(CPSBS). The samples were then annealed in a sealed evacuated quartz
tube at 550 $^{\circ}$C for 2 h and quenched by dropping the quartz tube
into cold water to activate the superconductivity.

The dc magnetic susceptibility was measured with a commercial SQUID
magnetometer (Quantum Design MPMS-1); the remnant field was removed with
the magnet reset procedure and the error in the applied field was less
than 0.01 mT. The resistivity $\rho_{xx}$ and the Hall resistivity
$\rho_{yx}$ were measured by using a standard six-probe method where the
contacts were made by attaching gold wires with a vacuum-cure silver
paint. The Hall coefficient $R_{\rm H}$ was calculated from the slope of
$\rho_{yx}(B)$. The specific heat $c_p$ was measured with a
relaxation-time method using a commercial equipment (Quantum Design
PPMS-9). To confirm the reproducibility of the specific-heat data, we
made detailed measurements on two superconducting samples with nearly
100\% shielding fractions, $x$ = 1.36 and 1.66, both of which had $T_c$
= 2.85 K.


\begin{flushleft} 
{\bf S1. Carrier density and Cu intercalation}
\end{flushleft} 

\begin{figure}[b]
\begin{center}
\includegraphics[clip,width=10cm]{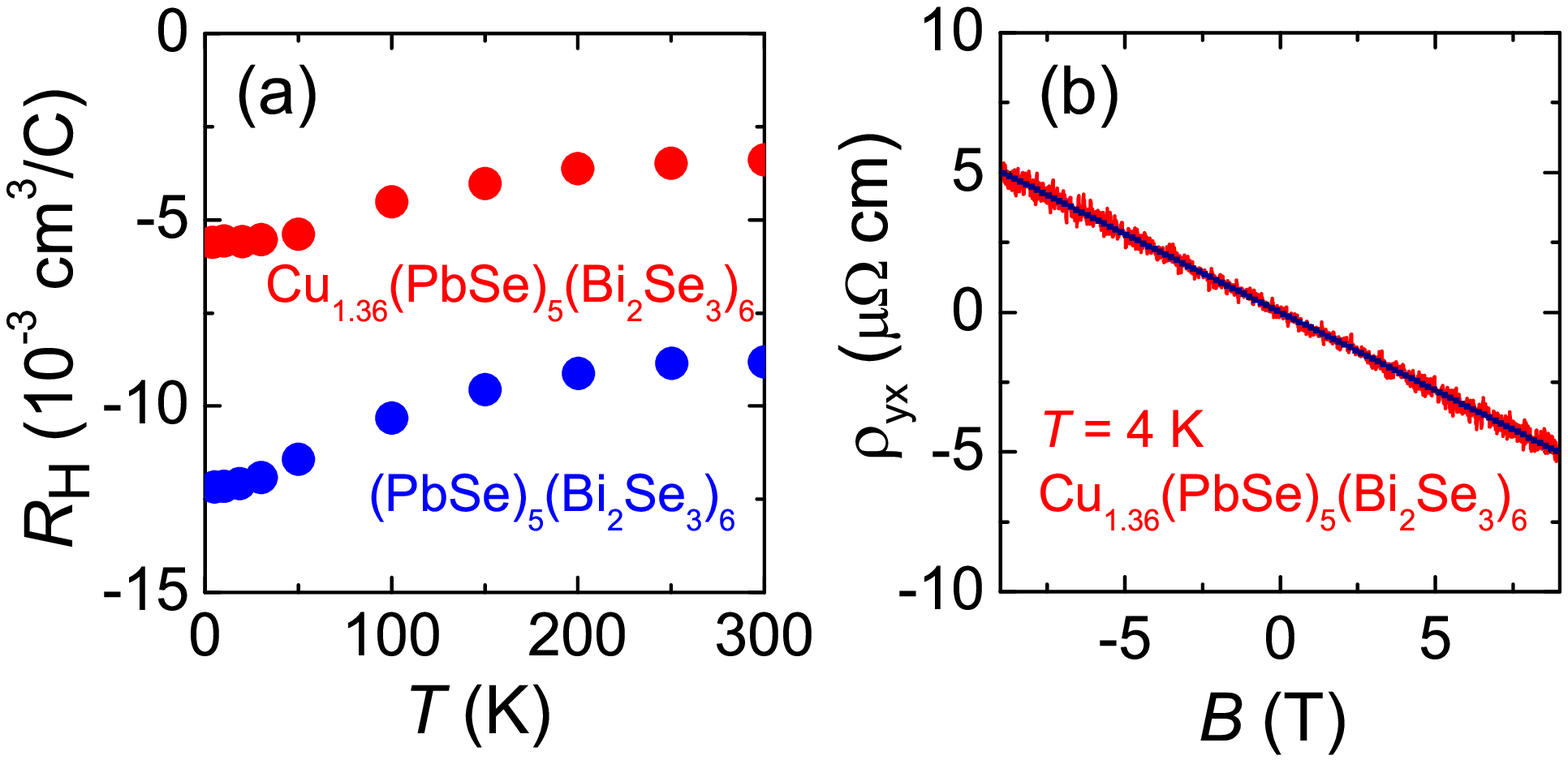}
\caption{
(a) Temperature dependencies of the Hall coefficient $R_{\rm H}$ in 
pristine PSBS and superconducting CPSBS ($x$ = 1.36). 
(b) Magnetic-field dependence of $\rho _{yx}$ measured in CPSBS 
($x$ = 1.36) at 4 K. The solid line emphasizes the linear nature
of the $B$ dependence.
}
\end{center}
\end{figure}

Temperature dependencies of $R_H$ in pristine PSBS and superconducting
CPSBS ($x$ = 1.36) are shown in Fig. S1(a). The carrier density $n_e$ is
calculated from the value of $R_H$ at 4 K, and it is $4 \times
10^{20}$ cm$^{-3}$ in PSBS and $1.2 \times 10^{21}$ cm$^{-3}$ in CPSBS.
The magnetic-field dependence of $\rho_{yx}$ in CPSBS at 4 K is
essentially linear, indicating that only one type of electron carriers
dominate the transport properties; namely, the physics is dominated by
only one band. This suggests that the topological and nontopological
bands of the Bi$_2$Se$_3$ unit observed in PSBS \cite{Nakayama} may
well have merged at the chemical potential of CPSBS that has been 
raised due to the electron doping.

In CPSBS samples, the volume density of Cu atoms is given by $n_{\rm Cu}
= x (d/M)N_A$, where $d$ = 7.715 g/cm$^3$ is the density of PSBS, $M$ =
5359.8 g/mol is the molar mass, and $N_A$ is the Avogadro constant. For
$x$ = 1.36, one obtains $n_{\rm Cu}$ = 1.18 $\times$ 10$^{21}$
cm$^{-3}$. On the other hand, the increase in the electron carrier
density upon Cu intercalation in this sample is given by $\Delta n_e$ =
$(1.2 \times 10^{21}) - (4 \times 10^{20})$ = $8 \times 10^{20}$
cm$^{-3}$. Therefore, one can estimate that each intercalated Cu
introduces $\Delta n_e/n_{\rm Cu}$ = 0.68 electron on average, which is
much more efficient than in Cu$_x$Bi$_2$Se$_3$ \cite{Hor}. 

\newpage
\begin{flushleft} 
{\bf S2. Specific heat analyses}
\end{flushleft} 

\begin{figure}
\begin{center}
\includegraphics[clip,width=7cm]{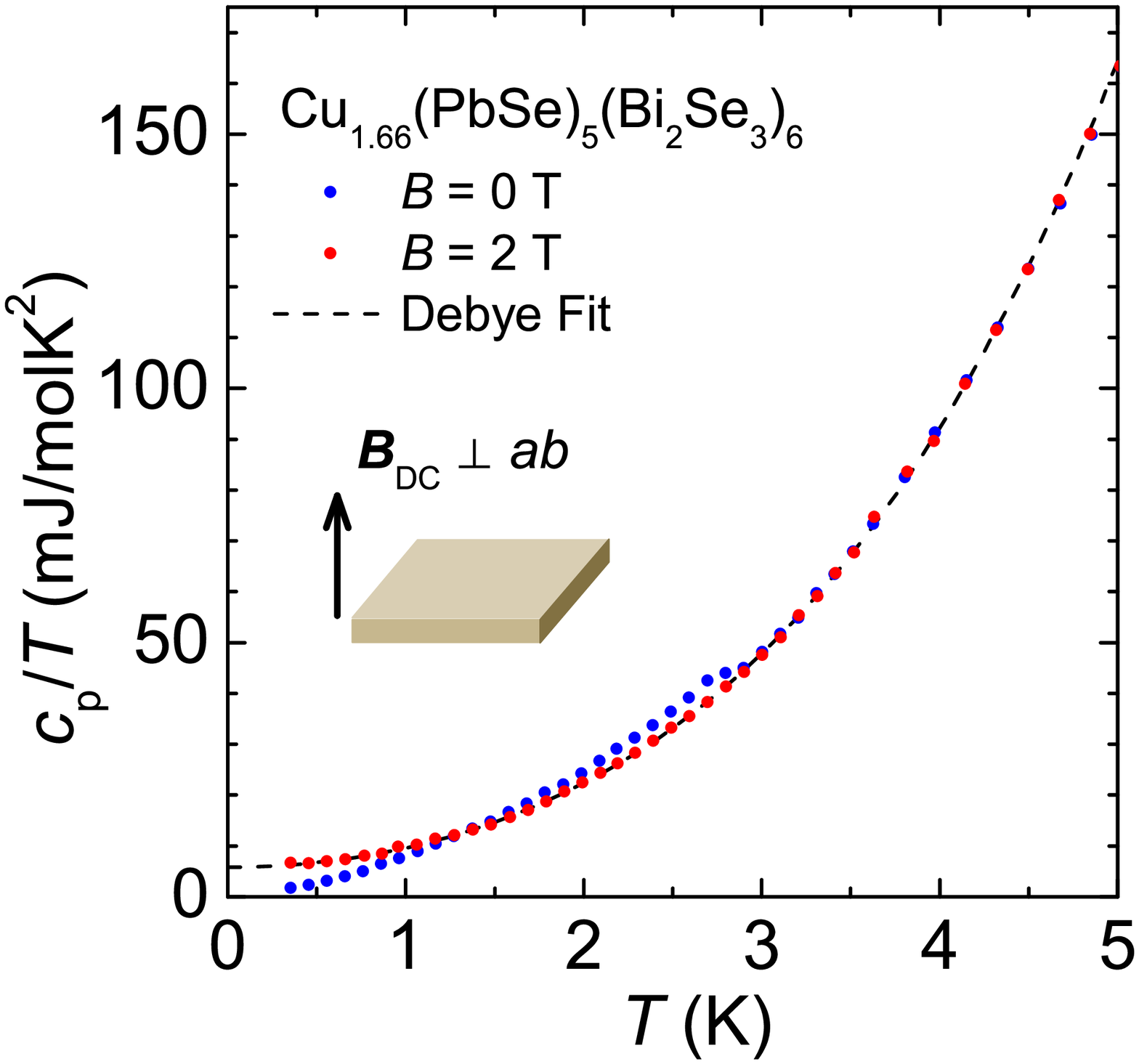}
\caption{$c_p/T$ vs $T$ data measured in 0 and 2 T applied perpendicular
to the $ab$ plane. The dashed line is the Debye fitting to the 2-T data.
}
\end{center}
\end{figure}

\begin{figure}[b]
\begin{center}
\includegraphics[clip,width=12cm]{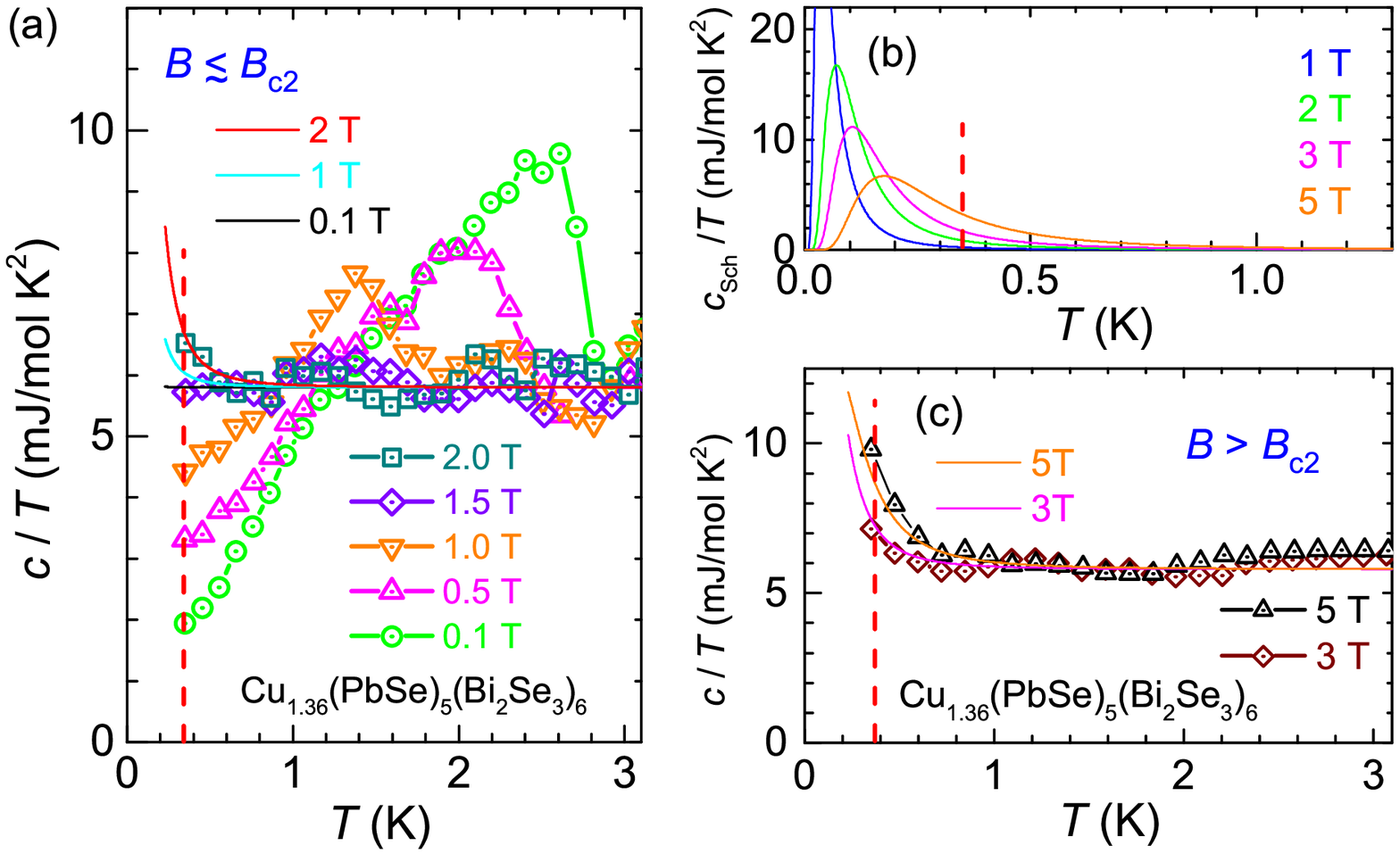}
\caption{
Schottky anomaly in the $x$ = 1.36 sample. 
(a) $(c_p - c_{\rm ph})/T$ vs $T$ data for $B \lesssim B_{c2}$ (symbols), 
together with the calculated $c_{\rm Sch}/T + \gamma_{\rm N}$ for 
$B$ = 0.1, 1, and 2 T (solid lines). 
(b) Theoretical curves of the expected Schottky contribution $c_{\rm Sch}/T$ 
coming from 0.037\% of $S$ = 1/2 moments with the Land\'e $g$ factor of 0.17. 
Those parameters are determined to consistently reproduce the data for the 
$x$ = 1.36 sample in 2 -- 5 T. (c) $(c_p - c_{\rm ph})/T$ vs $T$ data
for $B > B_{c2}$ and the calculated $c_{\rm Sch}/T + \gamma_{\rm N}$ for
$B$ = 3 and 5 T (solid lines). In panels (a)-(c), vertical red dashed lines 
mark the lowest experimental temperature. 
}
\end{center}
\end{figure}

The temperature dependence of the total specific heat $c_p$, which
includes both the phononic and electronic contributions, $c_{\rm ph}$
and $c_{\rm el}$, respectively, is shown in Fig. S2 for 0 and 2 T. The
conventional Debye fitting using 
\begin{equation}
c_p = c_{\rm el} + c_{\rm ph} = \gamma_{\rm N}T + A_3 T^3 + A_5 T^5
\end{equation}
to the 2 T data, which represent mostly the normal-state behavior in the
fitted temperature range, describes well the data up to 5 K. From this
fitting we obtain the normal-state electronic specific-heat coefficient
$\gamma_{\rm N}$ = 5.89 mJ/mol K$^2$. Assuming that the system is
quasi-2D, this $\gamma_{\rm N}$ corresponds to the effective mass
$m^*=(3\hbar^2 c^* \gamma_{\rm N})/(V_{\rm mol} k_{\rm B}^2)$ =
2.6$m_{\rm e}$, where $c^*$ = 5.06 nm is the lattice constant along the
$c^*$ direction, $V_{\rm mol}$ = 694.7 cm$^3$/mol is the molar volume,
and $m_{\rm e}$ is the free electron mass. The coefficients of the
phononic contribution are $A_3$ = 3.73 mJ/mol K$^4$ and $A_5$ = 0.10
mJ/mol K$^6$, and the former gives the Debye temperature $\theta_D$ =
153.1 K. 

We found that in high magnetic fields (above $\sim$1.5 T), a Schottky
anomaly becomes noticeable in the $c_p(T)$ data at low temperature. We
therefore analyzed the small Schottky contribution $c_{\rm Sch} \equiv
c_p - c_{\rm ph} - c_{\rm el}$ using the two-level Schottky model with
free $S$ = 1/2 moments \cite{Schottky, Moler},
\begin{equation}
c_{\rm Sch}(T,B) = \frac{n x^2 e^x}{(1 + e^x)^2}
\quad \quad \left(x \equiv \frac{g \mu _B B}{k_B T}\right),
\end{equation}
where $g$ is the Land\'{e} $g$ factor and $n$ is a coefficient in the
unit of the universal gas constant $R$. The temperature below which an
upturn starts is determined solely by the $g$ factor, which is found to
be 0.17 from the data of the $x$ = 1.36 sample in high magnetic fields
above $B_{c2}$ [Fig. S3(c)]; such a small $g$ factor has been reported to
come from an anisotropy caused by crystal fields \cite{Holmes, Hannak,
Jeune, Meulen} or from hyperfine-enhanced nuclear magnetic moments
\cite{Andres, Movshovich}. The upturns in the 2 -- 5 T data shown in
Figs. S3(a) and S3(c) are consistently reproduced by Eq. (2) with the
coefficient $n$ = 3.1 mJ/mol K, which corresponds to the free-moment
concentration of only $\sim$0.037\%. It is worth emphasizing that the
superconductivity in CPSBS is already suppressed above $\sim$2 T at 0.35
K, so the observed Schottky anomaly is largely irrelevant when one
discusses the specific-heat behavior in the superconducting state. The
calculated curves in Fig. S3(b) show that the peak due to this Schottky
anomaly is expected to occur at a much lower temperature than our
experimental range. 

\begin{figure}
\begin{center}
\includegraphics[clip,width=8.5cm]{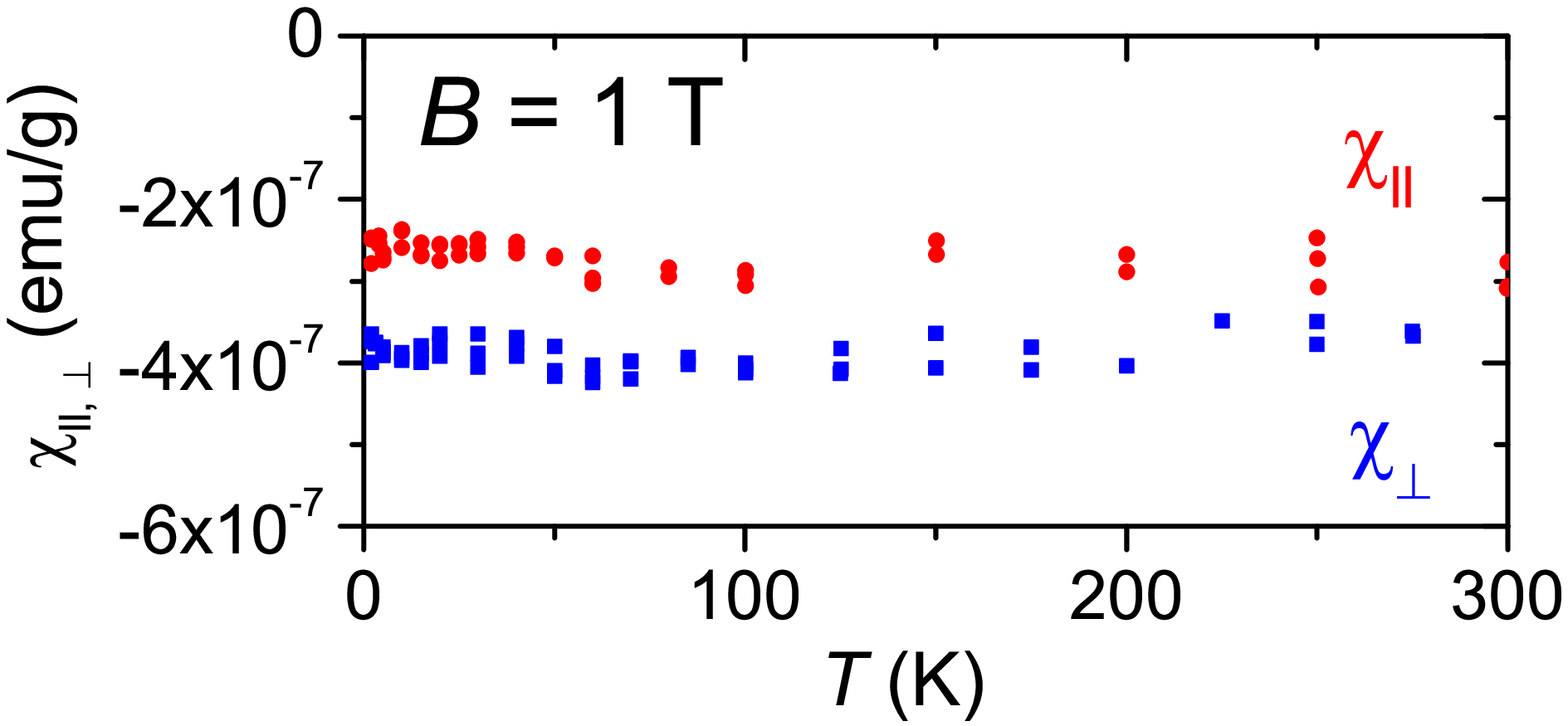}
\caption{
Normal-state magnetic susceptibilities $\chi_{\perp}$ and
$\chi_{\parallel}$ measured on the $x$ = 1.66 sample down to 1.8 K in 
1 T applied perpendicular and parallel to the $ab$ plane, respectively.}
\end{center}
\end{figure}

To gain insight into the origin of this small Schottky anomaly, we have
measured the normal-state magnetic susceptibilities $\chi_{\perp}$ and
$\chi_{\parallel}$ for $B$$\perp$$ab$ and $B$$\parallel$$ab$,
respectively, down to the lowest temperature of our SQUID magnetometer,
1.8 K. The result is shown in Fig. S4, where one can see that there is
no visible Curie behavior in neither of the field directions. If the
Schottky anomaly was due to $\sim$0.037\% of $S$ = 1/2 free electron
spins that happen to have a small $g$ factor for $B$$\perp$$ab$ due to
crystal fields, they should give rise to a visible Curie behavior for
$B$$\parallel$$ab$ above 1.8 K, but we did not observe it. Therefore,
one may conclude that the observed small Schottky anomaly is likely to
be due to hyperfine-enhanced nuclear magnetic moments, which originate
from a minority valence state or a minority isotope of the constituent
elements of CPSBS, or from some impurities that may have entered into
the samples during the Cu intercalation process. However, it is
difficult to name the actual element/isotope, because various nuclei can
have hyperfine-enhanced moments and the concentration of the nucleus in
question is only $\sim$0.037\%. In any case, the determination of the
exact source of the weak Schottky anomaly is not very important for the
present study, as long as its contribution can be duly subtracted.

\begin{figure}
\begin{center}
\includegraphics[clip,width=12cm]{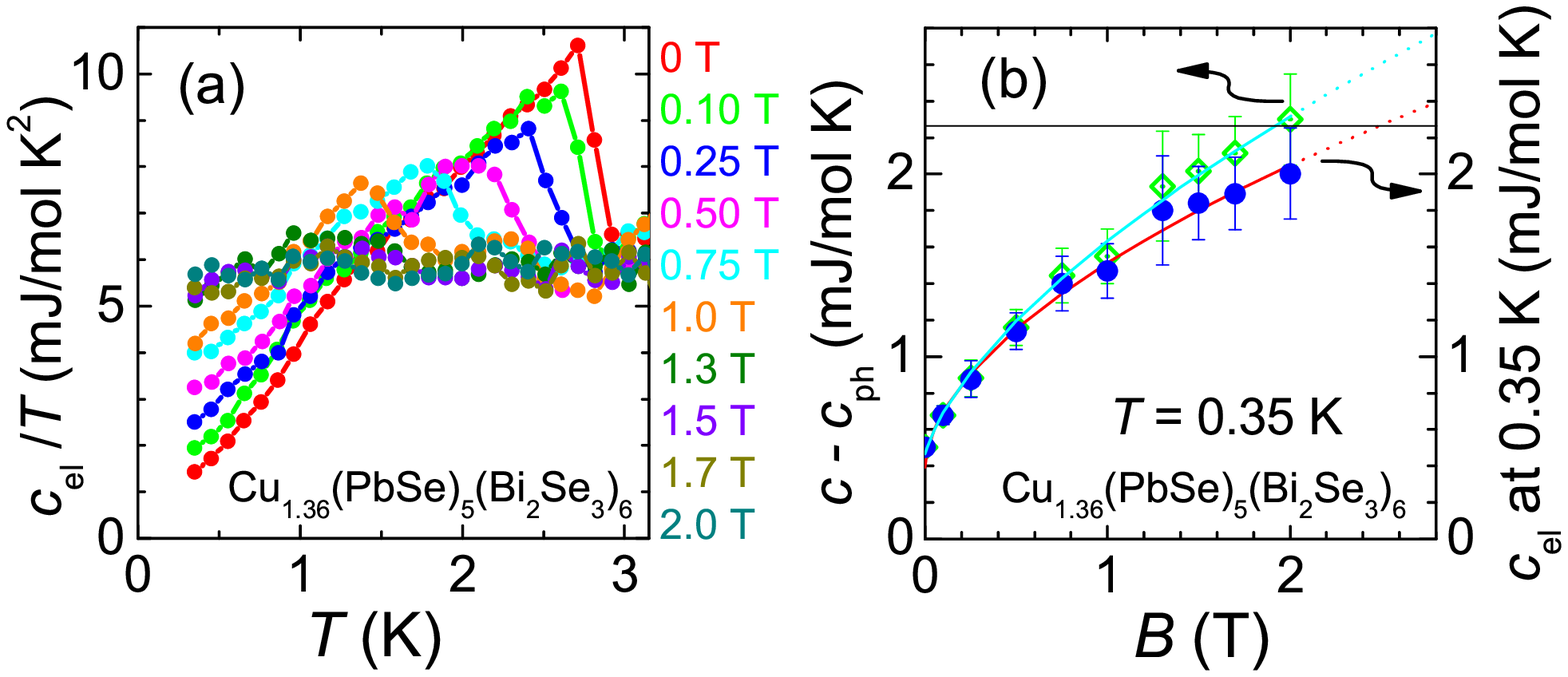}
\caption{
(a) $c_{\rm el}/T$ vs $T$ plots for various magnetic fields
($B$$\perp$$ab$) after subtracting the small Schottky contribution in
CPSBS ($x$ = 1.36).
(b) Magnetic-field dependence of $c_{\rm el}$ at 0.35 K taken from
the data shown in (a). The red solid line is the best fit of the
function $aB^n + c_0$ to the data and gives $n = 0.56 \pm 0.06$, $a =
1.12 \pm 0.06$ mJ mol$^{-1}$ K$^{-1}$ T$^{-n}$, and $c_0 = 0.39 \pm
0.05$ mJ mol$^{-1}$ K$^{-1}$. For comparison, we also show with
light-green diamonds the $B$ dependence of $c - c_{\rm ph}$ at 0.35 K
without subtracting the Schottky contribution; the light-blue solid line
is the best fit of the same function, yielding $n = 0.63 \pm 0.06$, $a =
1.3 \pm 0.1$ mJ mol$^{-1}$ K$^{-1}$ T$^{-n}$, and $c_0 = 0.4 \pm 0.5$
mJ mol$^{-1}$ K$^{-1}$. Horizontal solid line corresponds to
$\gamma_{\rm N}T$ at $T$ = 0.35 K.
}
\end{center}
\end{figure}

To make our best effort to quantify the magnetic-field dependence of
$c_{\rm el}$ in the superconducting state, we calculated
$c_{\rm Sch}(T,B)$ (which is only $\lesssim$ 20\% of $c_{\rm el}$ at 
0.35 K in 2 T and is negligible in 1 T)
and subtracted it from all the specific-heat data in
magnetic fields. The $c_{\rm el}/T$
vs $T$ data measured in the $x$ = 1.36 sample in various magnetic fields
applied perpendicular to the $ab$ plane are shown in Fig. S5(a) after the
subtraction of $c_{\rm Sch}(T,B)/T$; for comparison, Fig. S3(a) shows the 
data for $(c_p - c_{\rm ph})/T$ before subtraction of
$c_{\rm Sch}/T$, as well as the calculated Schottky component in terms
of $c_{\rm Sch}(T,B)/T + \gamma_{\rm N}$ for selected field values. 
The $c_{\rm el}(B)$ behavior at 0.35
K for this sample is shown in Fig. S5(b), which is best described with
$c_{\rm el} \sim B^{0.56}$; in this figure, we also show the uncorrected 
$c_{\rm el}(B)$ data including the Schottky contribution, the effect of 
which does not qualitatively change the behavior. The power 0.56 obtained 
for $c_{\rm el}(B)$ is consistent with the result for the $x$ = 1.66 
sample shown in the main text [Fig. 3(b)]. The parameters of the Schottky 
contribution used for obtaining the $c_{\rm el}/T$ data for the $x$ = 1.66 
sample shown in Fig. 3(a) of the main text were the same as those for the 
$x$ = 1.36 sample. 

As one can see in Fig. S5(a), the $c_{\rm el}/T$ vs $T$ behavior broadens
significantly in high magnetic fields above $\sim$1 T, which makes it
impossible to determine the mid-point of the specific-heat jump. We
therefore try to extract the information about $B_{c2}$ from $c_{\rm
el}/T$ by determining, with a certain error bar,  the onset temperature 
below which $c_{\rm el}/T$ deviates from $\gamma_{\rm N}$, and 
such data are plotted in Fig. 4(b) of the main text.

\begin{flushleft} 
{\bf S3. Superconducting parameters}
\end{flushleft} 

\begin{figure}[b]
\begin{center}
\includegraphics[clip,width=15.5cm]{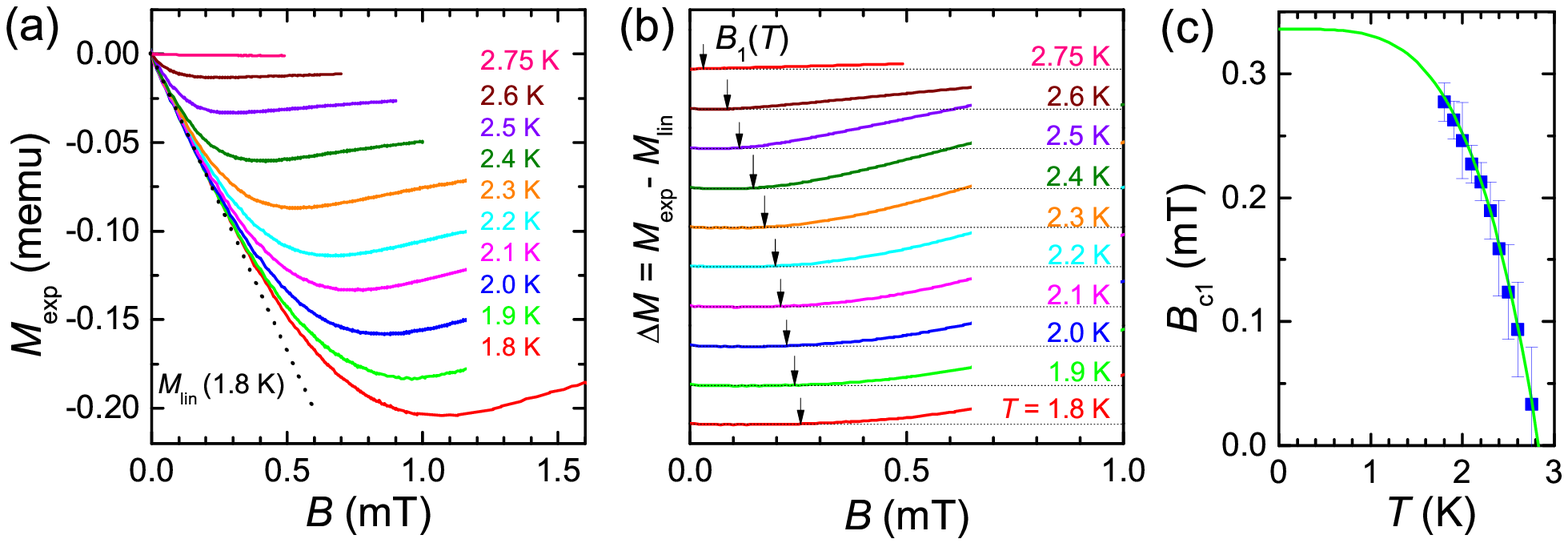}
\caption{
(a) Initial $M(B)$ behavior of the $x$ = 1.66 sample after 
zero-field cooling to various temperatures.
(b) Plots of $\Delta M \equiv M - aB$, where $a$ is the initial 
slope, together with the determination of $B_{\rm 1}$ shown by arrows.
(c) $B_{c1 \parallel}$ vs.\ $T$ phase diagram; the solid line is 
a fit to the empirical formula shown in the text.
}
\end{center}
\end{figure}

Figure S6 summarizes the results of the magnetization measurements of
the $x$ = 1.66 sample to determine the lower critical field $B_{c1}$.
Figure S6(a) shows $M(B)$ curves measured after zero-field cooling to
various temperatures in magnetic field applied parallel to the $ab$
plane. We define $B_{\rm 1}$ at each temperature as the value at which
the $M(B)$ data deviates from its initial linear behavior, as can be
seen in Fig. S6(b). To obtain $B_{c1 \parallel}$, those $B_{\rm 1}$ values
are corrected for the demagnetization effect, though it is small for
$B$$\parallel$$ab$: Using the approximation given for the slab geometry
\cite{brandt99a}, we obtain $B_{c1 \parallel}$ = $B_{\rm 1} / {\rm
tanh}\sqrt{0.36b/a}$, with the aspect ratio $b/a = 1.6/0.23$ in the
present case. The resulting $B_{c1 \parallel}$ values are shown in Fig.
S6(c). To determine the 0-K limit, we used the empirical formula $B_{\rm
c1}(T) = B_{\rm c1}(0)[1-(T/T_c)^4]$ \cite{Kriener12} and obtained
$B_{c1 \parallel}(0)$ = 0.34 mT. Note that the flux pinning in the
present system is weak as evidenced by the small magnetic hysteresis
[Fig. S7], which supports the reliability of the determination of
$B_{c1}$ using the above method \cite{Kriener11_1}.

\begin{figure}
\begin{center}
\includegraphics[clip,width=8cm]{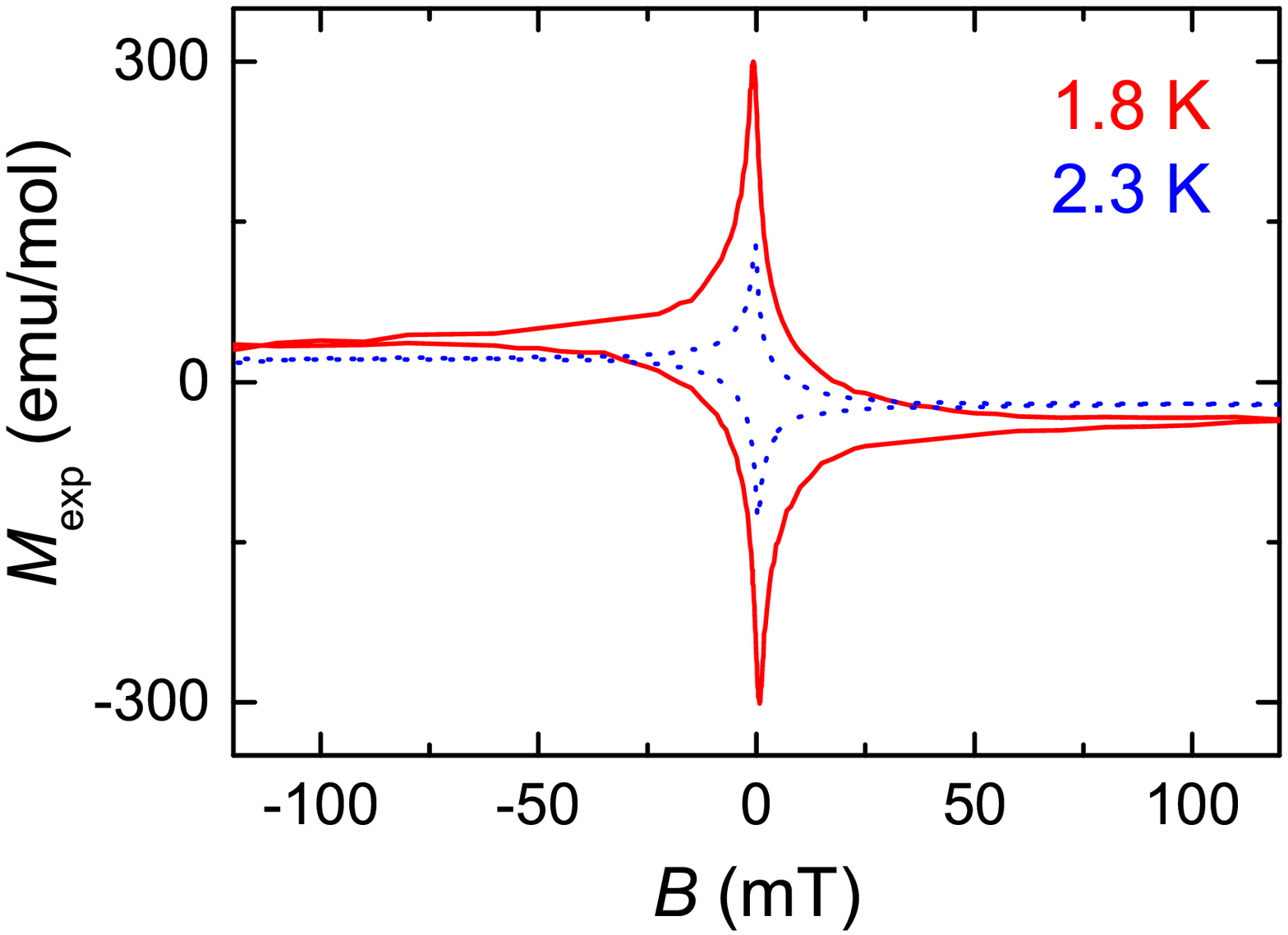}
\caption{
$M(B)$ curves measured on the $x$ = 1.66 sample at 1.8 and 2.3 K after 
subtracting the diamagnetic background.
}
\end{center}
\end{figure}

From $B_{c2 \perp}$ = 2.6 T, the coherence length $\xi_{ab} =
\sqrt{\Phi_0/(2\pi B_{c2 \perp})}$ = 11.3 nm is obtained, while from
$B_{c2 \parallel}$ = 4.3 T, we use $\xi_{ab}\xi_{c^*} = \Phi_0/(2\pi
B_{c2 \parallel})$ and obtain $\xi_{c^*}$ = 6.8 nm. The anisotropy ratio
is calculated as $\gamma = B_{c2 \parallel}/B_{c2 \perp} = \lambda_{c^*}
/\lambda_{ab}$ = 1.65; here, $\lambda_{c^*}$ and $\lambda_{ab}$ are the
penetration depths along the $c^*$ and $ab$ directions, respectively.
Since we have the $B_{\rm c1}$ value only for $B$$\parallel$$ab$, we
define the effective GL parameter $\kappa_{ab} \equiv \sqrt{\lambda_{ab}
\lambda_{c^*} / \xi_{ab} \xi_{c^*}}$ and use $B_{c1 \parallel} = \Phi_0
\ln{\kappa_{ab}} / (4\pi \lambda_{ab}\lambda_{c^*})$ together with
$B_{c2 \parallel}/B_{c1 \parallel} =
2\kappa_{ab}^2/(\ln{\kappa_{ab}}+0.5)$ \cite{Clem, Hu} to obtain
$\kappa_{ab} \approx$ 192. We then obtain the thermodynamic critical
field $B_{\rm c} = \sqrt{B_{c1 \parallel} B_{c2 \parallel} /
\ln{\kappa_{ab}}}$ = 16.6 mT and the penetration depths $\lambda_{ab} =
\kappa_{ab} \sqrt{\xi_{ab} \xi_{c^*}/\gamma}$ = 1.3 $\mu$m and
$\lambda_{c^*}$ = 2.2 $\mu$m.

\end{document}